\documentclass[
 reprint,
 amsmath,amssymb,
 aps,
]{revtex4-2}

\usepackage{graphicx}
\usepackage{dcolumn}
\usepackage{bm}
\usepackage{braket}
\usepackage{siunitx}
\usepackage{hyperref}
\hypersetup{colorlinks,citecolor=blue,linkcolor=blue,urlcolor=blue}

\begin{document}

\preprint{APS/123-QED}

\title{Quantum Optomagnetics in Graphene}

\author{Sina Abedi}
 \email{s3abedi@uwaterloo.ca}
 \affiliation{
  Department of Electrical and Computer Engineering,\\
  University of Waterloo, Waterloo, Ontario, N2L 3G1, Canada
 }
\author{A. Hamed Majedi}
 \altaffiliation[Also at ]{Perimeter Institute for Theoretical Physics, Waterloo,
 Ontario, Canada}
 \email{ahmajedi@uwaterloo.ca}
 \affiliation{Department of Electrical and Computer Engineering and Department of Physics and  Astronomy,\\
 Waterloo Institute for Nanotechnology, University of Waterloo, Waterloo, Ontario, N2L 3G1,  Canada}

\date{\today}

\begin{abstract}
Graphene can be magnetized through nonlinear response of its orbital angular momentum to an intense circularly polarized light. This optomagnetic effect can be well exemplified by the Inverse Faraday Effect (IFE) where an optically-generated DC magnetization leads to graphene's optical activity. We provide a single-particle quantum mechanical model of an IFE in graphene by solving Schr\"odinger's equation in the presence of a renormalized Hamiltonian near a Dirac point in the presence of circularly polarized monochromatic light. We derive an analytical expression for DC magnetization based on non-perturbative and dressed states of quasi-electrons where their energy spectrum is isotropically gapped by the circularly polarized light. Optical rotatory power is then computed through the gyroelectric birefringence where a measurable polarization rotation angle under moderate and intense optical radiations is predicted.
\end{abstract}

\maketitle

\section{Introduction}
Strong light-matter interaction in two-dimensional massless Dirac materials, i.e. graphene, leads to a new frontier at a merger of condensed matter physics and nonlinear and quantum optics~\cite{hendry2010coherent,koppens2011graphene,gu2012regenerative, semnani2016nonlinear}. Low-energy Dirac electronic band structure, topologically-protected band crossing and chirality of quasi-electrons in graphene provide a basis for a large number of peculiar properties such as strong nonlinear optical effect. Non-perturbative nonlinear optics of graphene has been proved to be a consequence of two major phenomena; singular nature of the interband dipole coupling~\cite{semnani2019anomalous,cheng2015third,vermeulen2022perspectives} and an electromagnetic field dressing leading to a renormalization of the energy spectrum~\cite{kristinsson2016control}. The former correlates with an exceptionally large and sign-alternating Kerr-type nonlinearity~\cite{vermeulen2018graphene,thakur2019experimental} and the latter associates with its gapped energy spectrum due to an intense circularly polarized electromagnetic field. Both phenomena can be used to investigate optomagnetism in graphene where DC and AC magnetization can be induced due to an intense not linearly polarized electromagnetic field \cite{kalashnikova2015ultrafast,battiato2014quantum}.

The hallmark of optomagnetics is the inverse Faraday effect where a not linearly polarized light can generate a DC magnetic field as a result of a second order nonlinear response of a material to the electric field~\cite{pitaevskii1961electric,kimel2004laser,majedi2021microwave}. The DC magnetization $\mathbf{M}_{\text{DC}}$ in its most basic form obeys Pitaevskii's relationship
\begin{align}
    \mathbf{M}_{\text{DC}}=i\gamma^{(1)}{\bf E}\times{\bf E}^*
    \label{Pitaevskii}
\end{align}
where $\gamma^{(1)}$ is the first order optical gyration coefficient and ${\bf E}$ is the complex electric field vector. The optical gyration coefficient has three significant properties. First, it does not possess any dependency on the crystallographic symmetry or electric and magnetic properties of the material hence it happens in all media. Secondly, materials with high values of magneto-optic coefficient or nonlinear optical susceptibility are more suitable to result in higher values of optical gyration coefficient. In fact, engineered materials and quantum confined nanostructures such as carbon nanotubes and quantum rings have been used to demonstrate the inverse Faraday effect and optically-induced persistent current with dissipationless electron transport \cite{kibis2021optically, koshelev2015resonant, kibis2011dissipationless}. Third, the optical gyration coefficient does not necessarily depend on absorption of photon or joule heating. The inverse Faraday effect is essentially a  non-resonant effect but can be enhanced through resonant structures.
Strong and non-perturbative nonlinear optical response of graphene due to its singular interband dipole coupling and strong band renormalization, on one hand, and its large magneto-optic and Kerr coefficient \cite{crassee2011giant, shimano2013quantum} on the other, make graphene a promising contender for optomagnetic effect suitable for applications ranging from nonlinear integrated photonic components to data storage and spintronics.

In this paper, we articulate a single-particle quantum mechanical model of the optomagnetic effect in graphene with a focus on IFE and the resulting optical rotatory power. Following the method presented by K. Kristinsson \textit{et al.}~\cite{kristinsson2016control} we use the renormalized wave functions of quasi-electrons in graphene under an intense circularly polarized electromagnetic wave considering energy spectrum modification, more importantly induced isotropic band gap formation around the Dirac point. Taking into account the hexagonal crystal structure of two Carbon atoms per unit cell, we find the Bloch wave functions and expectation value of angular momentum operator, obtaining the expression for a field-induced magnetization. We find a purely quantum mechanical, non-singular and non-perturbative DC magnetization of graphene in terms of the electric field strength, dressed frequency of the system and the modified band structure parameters of graphene. We show that our DC magnetization corroborates with the generalized Pitaevskii's relationship in a perturbative regime presented in~\cite{majedi2020nonlinear} and those are based on quasiclassical and linear quantum theory in a non-dressed field matter regime by I. D. Tokman {\it et. al.} ~\cite{tokman2020inverse}. We find the first order optical gyration coefficient of graphene and predict the optical rotatory power using the permittivity tensor. Measurable polarization rotation angle in a moderate and strong field intensity in graphene is predicted to be achieved in a conventional optical pump-probe experiment.
\section{Quantum Mechanical Formulation}
The electronic band structure of graphene is obtained through the tight-binding method.
Two Bloch functions are constructed from the atomic orbitals for the atoms
in sublattices A and B. These Bloch functions describing the $\pi$-states arising from $2 p_z$
orbitals provide the basis functions for the graphene Hamiltonian~\cite{saito1998physical,katsnelson2012graphene}. The special high symmetry points in the Brillouin zone are~\cite{neto2009electronic}
\begin{align}
    \mathbf{K} = \left(\frac{2 \pi}{3 a}, \frac{2 \pi}{3 \sqrt{3} a} \right), \mathbf{K}' = \left(\frac{2 \pi}{3 a}, -\frac{2 \pi}{3 \sqrt{3} a} \right), \mathbf{M} = \left(\frac{2 \pi}{3 a}, 0 \right)
    \label{eq:2}
\end{align}
The effective Hamiltonian near the Dirac point $K$ is written as
\begin{align}
    \hat{\mathcal{H}}_K = v \bm{\sigma} \cdot \hbar \mathbf{k}
    \label{eq:3}
\end{align}
where $\bm{\sigma} = (\sigma_x, \sigma_y)$ is the Pauli matrix vector, $v \approx c/300$ the Fermi velocity in graphene and $\mathbf{k}$ the wave vector of the electron.
Applying circularly polarized light to a sheet of graphene will modify the Hamiltonian in Eq.~\eqref{eq:3} with an extra term
\begin{align}
    \hat{\mathcal{H}} = v \bm{\sigma} \cdot (\hbar \mathbf{k} - e \mathbf{A})
    \label{eq:4}
\end{align}
where $\mathbf{A} = (A_x, A_y)$ is the magnetic vector potential, $\mathbf{k} = (k_x, k_y) = (k \cos{\varphi}, k \sin{\varphi})$ the wave vector and $\varphi$ the azimuth angle of the electron wave vector. The magnetic vector potential is expressed as
\begin{align}
    \mathbf{A} = (A_0 \cos{\omega t}, A_0 \sin{\omega t})
    \label{eq:5}
\end{align}
where $A_0 = \frac{E_0}{\omega}$. The Hamiltonian can be broken up into two parts $\hat{\mathcal{H}} = \hat{\mathcal{H}}_0 + \hat{\mathcal{H}}_k,$ where
\begin{align}
    \hat{\mathcal{H}}_0 & = (-v e A_0) \begin{pmatrix}
        0 & e^{i \omega t} \\
        e^{-i \omega t} & 0
    \end{pmatrix} 
    \label{eq:6} \\
    \hat{\mathcal{H}}_k & = \begin{pmatrix}
        0 & v \hbar (k_x - i k_y) \\
        v \hbar (k_x + i k_y) & 0
    \end{pmatrix}
    \label{eq:7}
\end{align}

Kristinsson \textit{et al.} (2016)~\cite{kristinsson2016control} showed that the wavefunction for the total Hamiltonian can be found by first solving the Schr\"odinger equation for the interaction Hamiltonian $i \hbar \frac{d}{dt} \psi_0 = \hat{\mathcal{H}}_0 \psi_0$, giving the following wavefunction
\begin{align}
    \psi_0^{\pm} &= \Bigg[\sqrt{\frac{\Omega \pm \hbar \omega}{2 \Omega}} \Phi_1^{'}(\mathbf{r}) e^{- \frac{i \omega t}{2}} \pm \sqrt{\frac{\Omega \mp \hbar \omega}{2 \Omega}} \Phi_2^{'}(\mathbf{r}) e^{\frac{i \omega t}{2}} \Bigg] \nonumber \\
    & \times \exp{\left(\pm \frac{i \Omega t}{2 \hbar} \right)}
    \label{eq:8}
\end{align}
where the new energy of the system dressed by circularly polarized light can be defined as $\Omega = \sqrt{(\hbar \omega)^2 + \left(\frac{2 v e E_0}{\omega} \right)^2}$. This dressed energy also gives the dressed frequency, i.e., $\omega_\text{dressed} = \sqrt{\omega^2 + \left(\frac{2 v e E_0}{\hbar \omega} \right)^2}$.
\begin{figure}
    \includegraphics[scale=0.5]{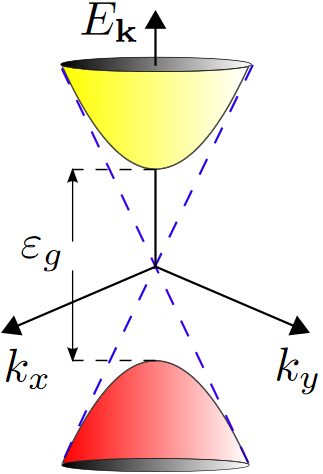}
    \caption{Isotropic energy band gap formation around the Dirac points, $\varepsilon_g$, under the application of an intense circularly polarized pump field.}
    \label{fig:graphene_gap}
\end{figure}

The total wavefunction $\psi_{\mathbf{k}}$ is then assumed to be a superposition of the two wavefunctions expressed in Eq.~\eqref{eq:8}, i.e., $\psi_\mathbf{k} = \xi^{+}(t) \psi_0^{+} + \xi^{-} \psi_0^{-}$. Solving Schr\"odinger's equation for the total Hamiltonian will lead to an expression for the total wavefunction
\begin{align}
    \psi_{\mathbf{k}} &= e^{-i \frac{\varepsilon_{\mathbf{k}}}{\hbar} t} \Bigg[\pm \sqrt{\frac{|\varepsilon_{\mathbf{k}}| \mp \frac{\varepsilon_g}{2}}{2 |\varepsilon_{\mathbf{k}}|}} e^{-i \frac{\varphi}{2}} \Bigg(\sqrt{\frac{\Omega + \hbar \omega}{2 \Omega}} \Phi_1^{'}(\mathbf{r}) \nonumber \\
    & + \sqrt{\frac{\Omega - \hbar \omega}{2 \Omega}} \Phi_2^{'}(\mathbf{r}) e^{i \omega t} \Bigg) \mp \sqrt{\frac{|\varepsilon_{\mathbf{k}}| \pm \frac{\varepsilon_g}{2}}{2 |\varepsilon_{\mathbf{k}}|}} e^{i \frac{\varphi}{2}} \nonumber \\
    & \times \Bigg(\sqrt{\frac{\Omega - \hbar \omega}{2 \Omega}} \Phi_1^{'}(\mathbf{r}) e^{-i \omega t} - \sqrt{\frac{\Omega + \hbar \omega}{2 \Omega}} \Phi_2^{'}(\mathbf{r}) \Bigg) \Bigg]
    \label{eq:9}
\end{align} 
where $\Phi_{1, 2}^{'}(\mathbf{r}) = \Phi_{1, 2}(\mathbf{r}) \varphi_{\mathbf{k}}(\mathbf{r})$ and $\Phi_{1, 2}(\mathbf{r})$ are the Bloch basis functions corresponding to the $2 p_z$ atomic orbital functions for the A and B sublattice atoms and $\varphi_{\mathbf{k}}(\mathbf{r}) = \frac{1}{\sqrt{S}} e^{i \mathbf{k} \cdot \mathbf{r}}$ with $S$ as the area of the graphene~\cite{kristinsson2016control}. As we are dealing with one unit cell in our subsequent calculations, we ignore the normalization factor $1/\sqrt{S}$. $\varepsilon_\mathbf{k}$ and $\varepsilon_g$ are the modified energy dispersion and the generated bandgap energy of graphene respectively after the application of circularly polarized light (see Fig.~\ref{fig:graphene_gap}):
\begin{align}
    \varepsilon_{\textbf{k}} &= \pm \sqrt{\left(\frac{\varepsilon_g}{2} \right)^2 + (\hbar v k)^2},
    \label{eq:10} \\
    \varepsilon_g &= \sqrt{(\hbar \omega)^2 + \left(2 v e A_0 \right)^2} - \hbar \omega
    \label{eq:11}
\end{align}
The energy gap can be expressed in terms of the difference between the dressed and the original frequency multiplied by $\hbar$. It is similar to the detuning for the generalized Rabi frequency of a two-level system.

\begin{figure}
    \includegraphics[width=0.42\textwidth]{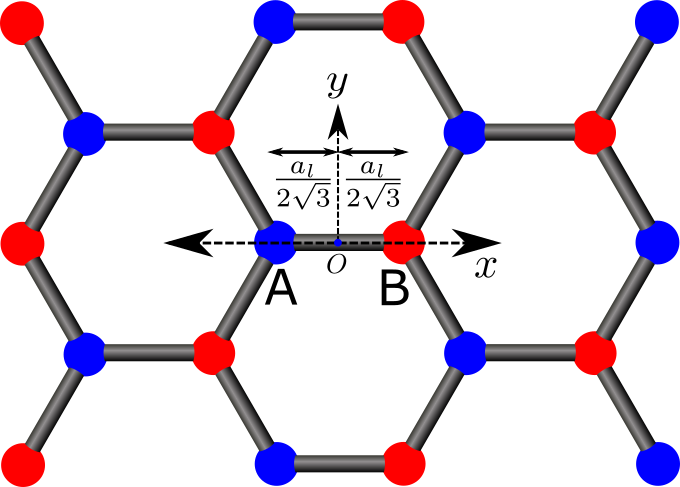}
    \caption{Coordinate reference frame for the shifts in the orbital functions $\Psi_{210}(\mathbf{r})$.}
    \label{fig:graphene_translation}
\end{figure}

\begin{figure}
    \includegraphics[width=0.449\textwidth]{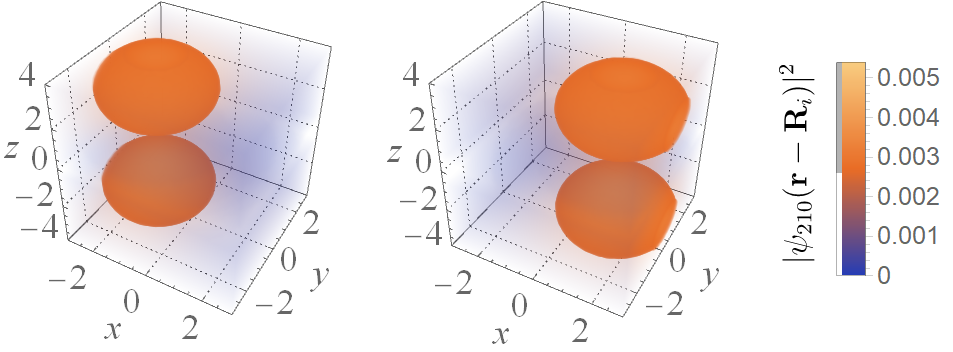}
    \caption{Orbital probability density for a shift in the direction of $\mathbf{R}_A$ and $\mathbf{R}_B$. Axes are in units of Bohr radius $a_0$.}
    \label{fig:graphene_orbital_shifts}
\end{figure}

As shown in Fig.~\ref{fig:graphene_translation}, these Bloch functions are constructed from the shift in the atomic orbital function $\Psi_{210}(\mathbf{r})$~\cite{saito1998physical,liu2018graphene}:
\begin{align}
    \Phi_1(\mathbf{r}) &= e^{i \mathbf{k} \cdot \mathbf{R}_A} \Psi_{210}(\mathbf{r} - \mathbf{R}_A) 
    \label{eq:12} \\
    \Phi_2(\mathbf{r}) &= e^{i \mathbf{k} \cdot \mathbf{R}_B} \Psi_{210}(\mathbf{r} - \mathbf{R}_B)
    \label{eq:13}
\end{align}
where $\mathbf{R}_A = -\frac{a_l}{2 \sqrt{3}} \mathbf{x}$, $\mathbf{R}_B = \frac{a_l}{2 \sqrt{3}} \mathbf{x}$ and $a_l \approx 2.46 \text{\AA}$ is the lattice constant of graphene. The $2 p_z$ atomic orbital for the carbon atom in graphene is expressed in Cartesian coordinates as (see Fig.~\ref{fig:graphene_orbital_shifts}):
\begin{align}
    \Psi_{210}(\mathbf{r}) = \frac{1}{\sqrt{32 \pi}} \left(\frac{6}{a_0} \right)^{5/2} z e^{-\frac{3}{a_0} \sqrt{x^2 + y^2 + z^2}}
    \label{eq:14}
\end{align}
where $a_0$ is the Bohr radius.

Magnetization can be expressed using the expectation of the total angular momentum operator
\begin{align}
    \braket{\hat{\mathbf{M}}} = \frac{N e}{2 m} \braket{\hat{\mathbf{L}}}
    \label{eq:15}
\end{align}
where $N$ is the atomic number density of the material ($3.82 \times 10^{19} \, \text{m}^{-2}$ for graphene), $e$ is the charge of an electron and $m$ is the mass of an electron. The total angular momentum operator $\hat{\mathbf{L}}$ is the sum of its mechanical and field angular momentum operators, denoted by $\hat{\mathbf{L}}_p$ and $\hat{\mathbf{L}}_A$, respectively, as 
\begin{align}
    \hat{\mathbf{L}} = (\mathbf{r} \times \mathbf{p}) - e \, (\mathbf{r} \times \mathbf{A}) = \hat{\mathbf{L}}_p + \hat{\mathbf{L}}_A
    \label{eq:16}
\end{align}
Since the expectation value of the mechanical angular momentum operator is zero, i.e. $ \braket{\hat{\mathbf{L}}_p}=0$, the expectation value of the angular momentum operator due to the applied electromagnetic field $\hat{\mathbf{L}}_A$ can be compactly written as
\begin{equation}
    \braket{\hat{\mathbf{L}}_A} =  B I_{11} + C I_{22}
    \label{eq:17}
\end{equation}
where $B$ and $C$ are expressions to be determined and
\begin{align}
    I_{ij} &= \int \Phi^*_i(\mathbf{r}) \varphi^*_{\mathbf{k}}(\mathbf{r}) \hat{\mathbf{L}}_{A} \Big[\Phi_j(\mathbf{r}) \varphi_{\mathbf{k}}(\mathbf{r}) \Big] \,d^3 \mathbf{r}
    \label{eq:18}
\end{align}
Performing the integrals in Eq.~\eqref{eq:18} yields the following expression
\begin{align}
    \mathbf{I} = -\mathbf{z} \sigma_z \frac{a}{2} e A_0 \sin{\omega t}
    \label{eq:19}
\end{align}
where $a \approx 1.42 \, \text{\AA}$ is the distance between nearest-neighbour atoms. Eq.~\eqref{eq:19} shows that the resulting magnetization will be purely in the $z$-direction
\begin{align}
    \braket{\hat{M}}_z &= \frac{N e^2 a \hbar \, \varepsilon_g E_0}{8 m |\varepsilon_\mathbf{k}| \Omega} \sin{\omega t} + \frac{N e^3 a \hbar v^2 k E_0^2}{4 m \omega^2 |\varepsilon_\mathbf{k}| \Omega} \nonumber \\
    & \times \Big[\sin{\varphi} + \sin{(2 \omega t - \varphi)} \Big]
    \label{eq:20}
\end{align}
Eq.~\eqref{eq:20} contains both the AC and DC components of the resulting magnetization. The DC component of the magnetization will be pointing in the opposite direction of the propagation of the circularly polarized light, i.e., in the $-z$-direction
\begin{align}
    \langle \hat{M} \rangle_{\text{DC}} = \frac{N e^3 a \hbar v^2 k_y E_0^2}{4 m \omega^2 |\varepsilon_\mathbf{k}| \Omega} 
    \label{eq:21}
\end{align}
Eq.~\eqref{eq:21} can be rewritten in terms of the bandgap energy $\varepsilon_g$ as
\begin{align}
    \langle \hat{M} \rangle_{\text{DC}} = \frac{ N e \hbar k_y a}{16 m} \times \frac{(\varepsilon_g + \hbar \omega)^2 - (\hbar \omega)^2}{(\varepsilon_g + \hbar \omega) \sqrt{(\varepsilon_g/2)^2 + (\hbar v k)^2}}
    \label{eq:22}
\end{align}
This form of the expression clearly shows the direct dependency of the DC magnetization on the generated bandgap in graphene due to the application of the circularly polarized pump-light. In the case where there is no bandgap, i.e., $\varepsilon_g = 0$, DC magnetization also vanishes since the numerator $(\varepsilon_g + \hbar \omega)^2 - (\hbar \omega)^2$ will become zero.

Expanding Eq.~\eqref{eq:21} as a Taylor series up to the sixth order in terms of $E_0$ as the variable, we will have
\begin{align}
    \langle \hat{M} \rangle_{\text{DC}} &= \frac{N e^3 a v k_y}{4 m \omega^3 \hbar k} E_0^2 - \frac{N e^5 a v^3 k_y}{2 m \omega^7 \hbar^3 k} E_0^4 \nonumber \\
    & + \frac{N e^7 a v^3 k_y (\omega^2 - 12 k^2 v^2)}{8 m \omega^{11} \hbar^5 k^3} E_0^6 + \dots
    \label{eq:23}
\end{align}
The orders of the Taylor polynomial will be $2, 4, 6, \dots$, thus the DC magnetization is a function of $E_0$ in the form of $|\mathbf{E_p \times E_p^{*}}|^{n}$ (where $n = 1, 2, 3, \dots$), as predicted in ~\cite{majedi2020nonlinear}. The above expression allows comparison between our result and similar studies that have expressed magnetization as a function of $E_0^2$, by taking the first term
\begin{align}
    \braket{\hat{M}}_{\text{DC}} \simeq \frac{N e^3 a v k_y}{4 m \omega^3 \hbar k} E_0^2
    \label{eq:24}
\end{align}
Comparison with Hertel~\cite{hertel2006theory,hertel2015macroscopic} and Tokman \textit{et al.} (2020)~\cite{tokman2020inverse} shows the same proportionality of the magnetization with the frequency of the applied light $\omega$, i.e., $\mathbf{M}_\text{DC} \propto \omega^{-3}$. Comparing Eq.~\eqref{eq:24} with Eq.~\eqref{Pitaevskii} the first order optical gyration coefficient will be
\begin{align}
    \gamma^{(1)} = \frac{N e^3 a v k_y}{8 m \omega^3 \hbar k}
    \label{eq:25}
\end{align}

However, the expressions above for the magnetization and gyration coefficients are only valid for one wave vector $\mathbf{k}$. Since the Hamiltonian in Eq.~\eqref{eq:4} was constructed for the regions in close proximity of the Dirac point $K$, all wave vectors $\mathbf{k}$ must be located in that same region. Therefore, magnetization in Eq.~\eqref{eq:21} must be integrated over all $\mathbf{k}$ values in the vicinity of the $K$ and $K'$ points to give
\begin{align}
    \langle \hat{M} \rangle_{\text{DC}} &= \frac{N e^3 a \hbar v^2 E_0^2}{2 m \omega^2 \Omega} \times \frac{\iint_{K,K'}\frac{dk_y \,dk_x}{\sqrt{(\varepsilon_g/2)^2 + (\hbar v)^2 (k_x^2 + k_y^2)}}}{\iint_{K,K'}\,dk_x}
    \label{eq:26}
\end{align}
We take a region 5 degrees on both sides of each of the $K$ and $K'$ points from the center of the reciprocal lattice, introduce $dk_x$ for integration over $\mathbf{k}$ and divide by the integral of $dk_x$.

\begin{figure}
    \includegraphics[width=0.45\textwidth]{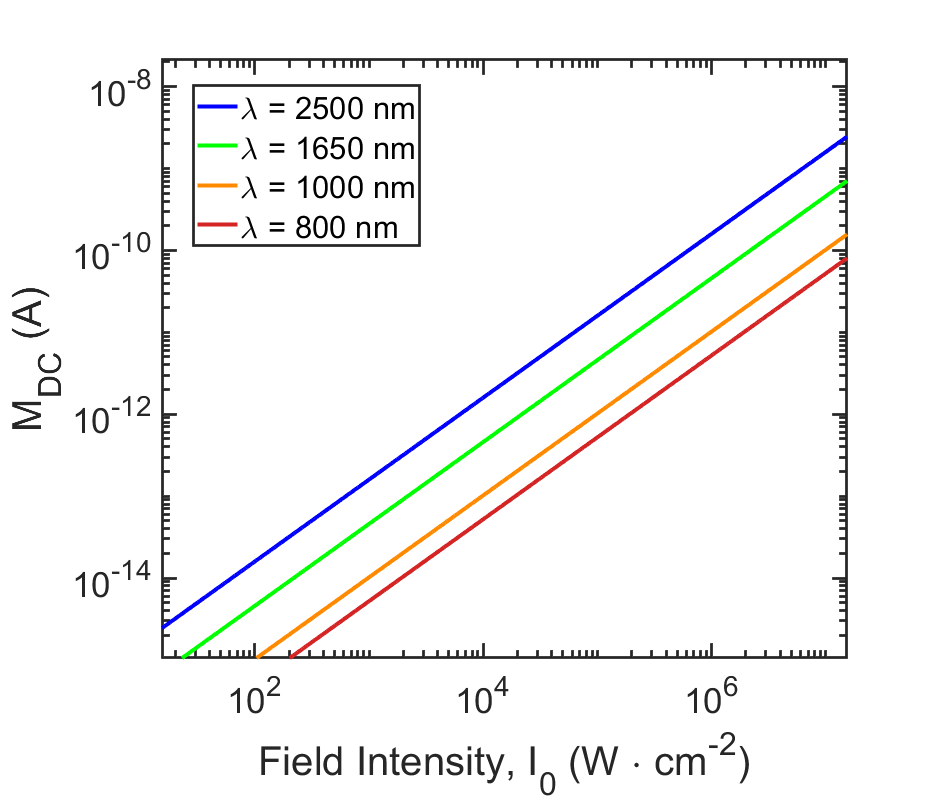}
    \caption{Magnitude of DC magnetization with respect to the pump-light intensity for wavelengths of $\SI{800}{\nm}$, $\SI{1650}{\nm}$, $\SI{1650}{\nm}$, $\SI{2500}{\nm}$. Magnetization saturates at a value of $6.4 \times 10^{-5} \, \si{\ampere}$ for all wavelengths after an intensity of approximately $10^{15} \, \si{\watt / \cm^2}$.}
    \label{fig:magnetization}
\end{figure}

Fig.~\ref{fig:magnetization} shows the magnetization with respect to applied laser intensities for wavelengths of $\SI{800}{\nm}$, $\SI{1000}{\nm}$, $\SI{1650}{\nm}$ and $\SI{2500}{\nm}$. Due to the effects of optical saturation, we have limited the range of applied intensities to a maximum of $10^7 \, \si{\watt / \cm^2}$. Semnani \emph{et al.} (2019) \cite{Semnani_2019} have shown the optical saturation in graphene for arbitrarily weak magnetic fields by modelling the electron dynamics using Semiconductor Bloch equations and using a many-body approach for the dynamics of carrier relaxation. They show that both the saturation region due to zero-detuning and the one due to strong interband coupling near the Dirac points are above $10^7 \, \si{\watt / \cm^2}$.

\begin{figure}
    \includegraphics[width=0.45\textwidth]{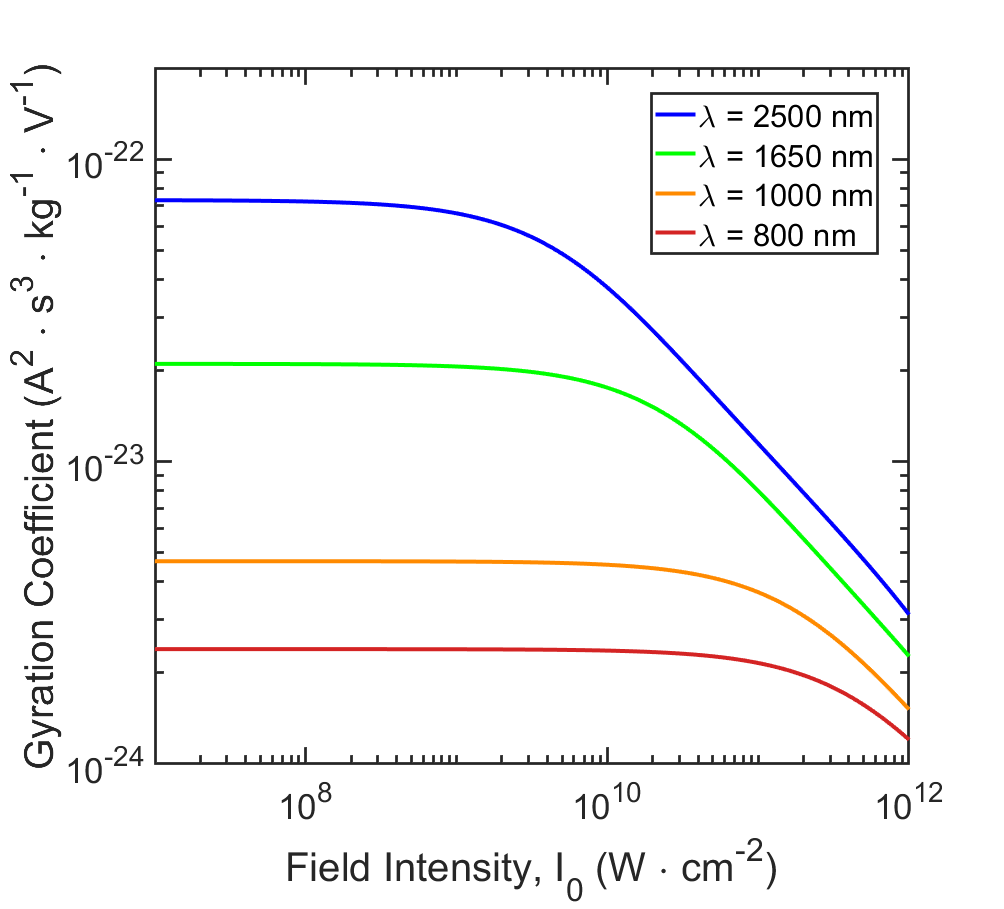}
    \caption{Gyration coefficient $\gamma$ with respect to the pump-light intensity for wavelengths of $\SI{800}{\nm}$, $\SI{1000}{\nm}$, $\SI{1650}{\nm}$ and $\SI{2500}{\nm}$.}
    \label{fig:gyration_coefficient}
\end{figure}

Fig.~\ref{fig:gyration_coefficient} shows the plot of the magnitude of the unperturbed gyration coefficient. This unperturbed gyration coefficient is approximately equal to the first order coefficient $\gamma^{(1)}$ for intensities below $\sim 10^{10} \; \si{\watt} / \si{\cm}^{2}$.

\begin{figure}
    \includegraphics[width=0.45\textwidth]{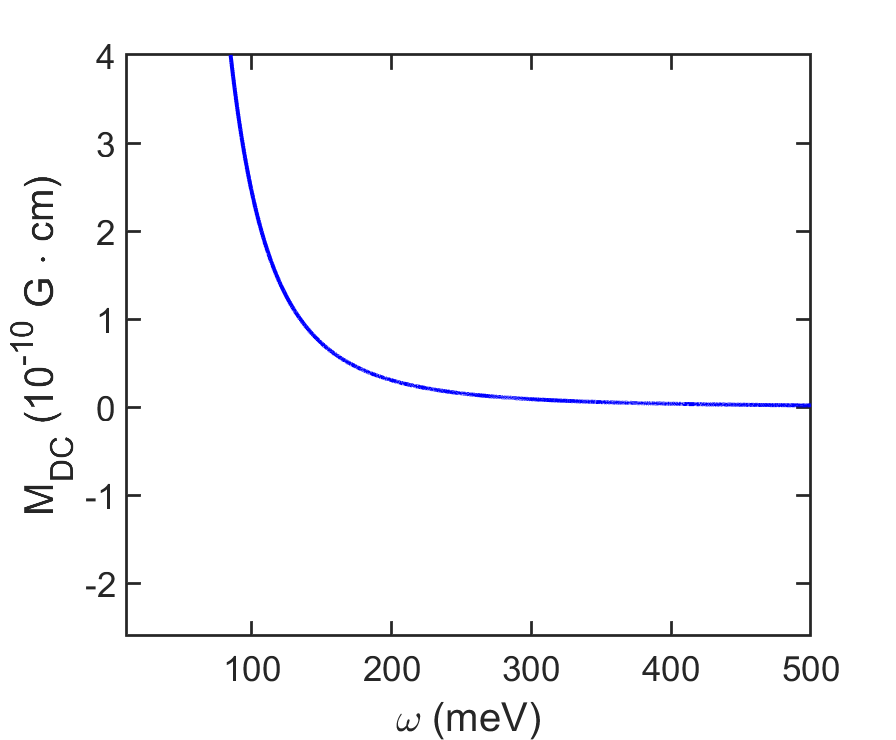}
    \caption{Magnetization with respect to the angular frequency $\omega$ of the circularly polarized pump-field for a constant field intensity $I_0 = \SI{10}{\kilo \watt / \cm^2}$. Compare to Fig.~(3) in \cite{tokman2020inverse}.}
    \label{fig:magnetization_frequency_dependence}
\end{figure}

Our results for the magnetization values show further similarities with those of Tokman \textit{et al.}~\cite{tokman2020inverse}. Fig.~\ref{fig:magnetization_frequency_dependence} displays a very similar trend to the one presented in \cite{tokman2020inverse}, except the values at, and in the approximate vicinity of $\hbar \omega = \SI{400}{\milli \electronvolt}$. Since our method does not introduce singularities into the magnetization expression, we see a continuous, decreasing trend instead of an asymptotic behaviour at $\SI{400}{\milli \electronvolt}$. The values obtained from Eq.~\eqref{eq:26} are on the same order as the ones presented in \cite{tokman2020inverse}, where an applied circularly polarized light with intensity of $\SI{10}{\kilo \watt / \cm^2}$ is used with varying frequencies. At the frequency corresponding to $\hbar \omega = \SI{100}{\milli \electronvolt}$, Tokman \textit{et al.}~\cite{tokman2020inverse} predict a magnetization of approximately $1.18 \times 10^{-10} \, \text{G} \cdot \text{cm}$, while the magnetization we obtain from Fig.~\ref{fig:magnetization_frequency_dependence} is a nearly identical $2.44 \times 10^{-10} \, \si{G \cdot \cm}$.
\section{Optical Rotatory Power}
We consider the co-propagation of a weak linearly polarized optical signal with frequency $\omega_s$ with an intense circularly polarized pump-light in graphene and a photo-induced magnetic field resulting in an optical birefringence. The magnetic field $B_0$ induced by the pump-light will be proportional to the DC magnetization as $B_0 = \mu_0 M_\text{DC}$. By adding this expression of the DC magnetic field inside the forcing function of the Drude-Lorentz model, we obtain a relative permittivity tensor at signal frequency $\omega_s$ which can be expressed in the following matrix form~\cite{majedi2020nonlinear}
\begin{align}
    \overline{\overline{\epsilon}}_r = \begin{pmatrix}
        1 + \frac{\omega_{pl}^2}{D_F^2(\omega_s)} D(\omega_s) & -i \omega_{pl}^2 \omega_c \frac{\omega_s}{D_F^2(\omega_s)} & 0 \\
        i \omega_{pl}^2 \omega_c \frac{\omega_s}{D_F^2(\omega_s)} & 1 + \frac{\omega_{pl}^2}{D_F^2(\omega_s)} D(\omega_s) & 0 \\
        0 & 0 & 1 + \frac{\omega_{pl}^2}{D(\omega_s)}
    \end{pmatrix}
    \label{eq:27}
\end{align}
we define~\footnote{Eq.~(35) in Ref.~\cite{majedi2020nonlinear} should only read as shown here in Eq.~\eqref{eq:28}.}
\begin{align}
    D_F^2(\omega_s) \triangleq D^2(\omega_s) - \omega_c^2 \omega_s^2
    \label{eq:28}
\end{align}
as the modified dispersion function due to the presence of the magnetic field through the cyclotron frequency $\omega_c = \frac{e B_0}{m} = \frac{\mu_0 e}{m} M_\text{DC}$, where
\begin{align}
    D(\omega_s) \triangleq \omega_0^2 - \omega_s^2 + i \omega_s \Gamma
    \label{eq:29}
\end{align}
is the dispersion function of a damped harmonic oscillator and $\Gamma$ is the friction term representing energy loss associated with the material absorption. $\omega_{pl} \triangleq \sqrt{\frac{n_{2D} e^2}{m d_\text{gr} \epsilon_0}}$ is the plasma frequency of graphene, $n_{2D}$ is the two-dimensional electron density of graphene and $d_\text{gr}$ is the effective thickness of graphene. Since graphene is a two-dimensional material, this effective thickness must be used to match units for the plasma frequency and magnetization (we set $d_\text{gr} \approx 1 \; \text{\AA}$ for our calculations.)~\cite{hendry2010coherent}. We obtain the relevant real and complex parts of the permittivity elements from the permittivity tensor in Eq.~\eqref{eq:27} for the complex propagation constants for an optical signal~\cite{majedi2020nonlinear}
\begin{align}
    \epsilon'_r &= \epsilon'_{xx} = \epsilon'_{yy} \approx 1 + \frac{\omega^2_{pl}}{|D(\omega_s)|^2} (\omega_0^2 - \omega_s^2) 
    \label{eq:39} \\
    \epsilon''_{xy} &= -\epsilon''_{yx} \approx \omega^2_{pl} \frac{(\omega_0^2 - \omega_s^2)^2 - \omega_s^2 \Gamma^2}{|D(\omega_s)|^4} \omega_s \omega_c
    \label{eq:30}
\end{align}
and
\begin{align}
    \beta_{1, 2} = \frac{2 \pi}{\lambda_0} \sqrt{\epsilon'_r} \Bigg(1 \mp \frac{\epsilon''_{xy}}{\epsilon'_r} \Bigg)
    \label{eq:32}
\end{align}
where $\beta_{1, 2}$ are the propagation constants and $\lambda_0$ is the wavelength of the pump-light.
\begin{figure}
    \includegraphics[width=0.45\textwidth]{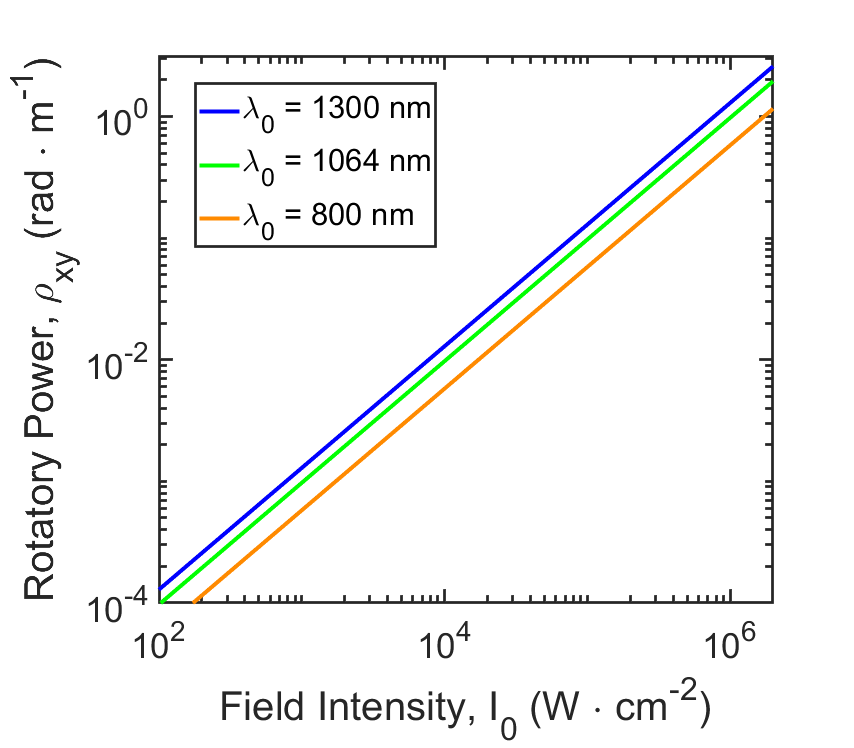}
    \caption{Rotatory power with respect to the pump-light intensity for wavelengths of $\SI{800}{\nm}$, $\SI{1064}{\nm}$ and $\SI{1300}{\nm}$ with the probe-light wavelength $\lambda_s =  \SI{1550}{\nm}$.}
    \label{fig:rotatory_power}
\end{figure}
The inverse Faraday effect can be quantified by the rotatory power $\rho_{xy}$, as the rotation angle per unit length
\begin{align}
    \rho_{xy} &= \frac{\beta_1 - \beta_2}{2} \approx -\frac{\pi}{\lambda_0} \frac{\epsilon''_{xy}}{\sqrt{\epsilon'_r}} \nonumber \\
    & \approx -\frac{N \pi n_{2D}  e^6 v^2 \hbar \omega_s E_0^2}{12 \sqrt{3} \, m^3 c^3 \epsilon_0^2 \omega_0 d_\text{gr}^2|\varepsilon_\mathbf{k}| \Omega} \nonumber \\
    & \times \frac{\left(\omega_0^2 - \omega_s^2 \right)^2 - \omega_s^2 \Gamma^2}{(\omega_0^2 - \omega_s^2)^4 + 2 (\omega_0^2 - \omega_s^2)^2 (\omega_s \Gamma)^2 + (\omega_s \Gamma)^4} \nonumber \nonumber \\
    & \times \frac{1}{\sqrt{1 + \frac{\omega^2_{pl}}{|D(\omega_s)|^2} (\omega_0^2 - \omega_s^2)}}
    \label{eq:33}
\end{align}
Fig.~\ref{fig:rotatory_power} shows the resulting rotation of a weak, linearly polarized light as a function of the intensity of the pump-light as it travels through the graphene layer. We observe  significant rotations from pump intensities starting from $\sim 10^5 \; \si{\watt} / \si{\cm}^{2}$. A conventional pump-probe optical experiment can reveal the predicted optical rotatory power in graphene under a moderate to intense pump field.
\section{Conclusions}
In conclusion, we have provided a quantum mechanical, non-perturbative and non-singular model of optomagnetic effect in graphene to describe the inverse Faraday effect. We presented analytical expression for photo-induced DC magnetization in terms of the electric field strength of the applied pump-light, the dressed energy of the system and band gap opening through the modified dispersion relation and quasi-electrons' wavefunction upon application of circularly polarized light. By expanding our non-perturbative magnetization as a perturbation of the electric field strength and comparing it with Pitaevskii's equation, we derived an expression for the first-order optical gyration coefficient which corresponds to the optical gyration coefficient in the perturbative approach. Lastly, we used the Drude-Lorentz model and the resulting relative permittivity tensor to derive the optical rotatory power that is measurable on a conventional optical pump-probe setup. Optomagnetic effects thus open a new possibilities to optically control the electronic and magnetic properties of graphene for applications in classical and quantum nonlinear optical devices.         

\begin{acknowledgments}
Authors acknowledge the financial support provided by Natural Sciences and Engineering Research Council of Canada (NSERC). S. A. acknowledges the financial support of Mitacs Globalink Research Internship, Canada.  
\end{acknowledgments}

\bibliographystyle{apsrev4-2}
\bibliography{quantum_optomagnetics_graphene}

\end{document}